\documentclass[a4paper,aps,pre,floatfix,nofootinbib,preprint,11pt]{revtex4}
\usepackage[dvips]{epsfig}
\usepackage[dvips]{graphicx}
\usepackage{latexsym,amsmath,verbatim}
\usepackage[usenames,dvipsnames]{color}
\newcommand{\be}{\begin{equation}}
\newcommand{\ee}{\end{equation}}

\usepackage{array}
\usepackage{amssymb,amsfonts,amsmath}

\begin{document}


\title{Griffiths phases and the stretching of criticality in brain networks}

\author{Paolo Moretti and Miguel A. Mu\~noz} 
\affiliation{Departamento de Electromagnetismo y
  F\'\i sica de la Materia and Instituto Carlos I de F\'\i sica
  Te\'orica y Computacional, Facultad de Ciencias, Universidad de
  Granada, 18071 Granada, Spain \\
  Correspondence and material requests should be
  addressed to M.A.M.  (email: {\tt mamunoz@onsager.ugr.es}). }


\begin{abstract} 
  Hallmarks of criticality, such as power-laws and scale invariance,
  have been empirically found in cortical networks and it has been
  conjectured that operating at criticality entails functional
  advantages, such as optimal computational capabilities, memory, and
  large dynamical ranges.  As critical behavior requires a high degree
  of fine tuning to emerge, some type of self-tuning mechanism needs
  to be invoked. Here we show that, taking into account the complex
  hierarchical-modular architecture of cortical networks, the singular
  critical point is replaced by an extended critical-like region which
  corresponds --in the jargon of statistical mechanics-- to a
  Griffiths phase.  Using computational and analytical approaches, we
  find Griffiths phases in synthetic hierarchical networks and also in
  empirical brain networks such as the human connectome and the {\it
  caenorhabditis elegans} one.  
  Stretched critical regions, stemming from structural
  disorder, yield enhanced functionality in a generic way,
  facilitating the task of self-organizing, adaptive, and evolutionary
  mechanisms selecting for criticality.
 \end{abstract}



\maketitle

Empirical evidence that living systems can operate near critical
points is flowering in contexts ranging from gene expression patterns
\cite{Kauffman08}, to optimal cell growth \cite{Kaneko}, bacterial
clustering \cite{Chen-bacteria}, or flocks of birds \cite{Cavagna12}.
In the context of neuroscience, synchronization patterns have been
shown to exhibit broadband criticality \cite{broadband}, critical
avalanches of spontaneous neural activity have been consistently found
both {\it in vitro} \cite{BP03,Plenz07,Beggs08} and {\it in vivo}
\cite{Peterman09}, and results from large-scale brain models based on
the human connectome show that only at criticality the brain structure
is able to support the dynamics seen in functional magnetic resonance
imaging (fMRI) recordings \cite{Haimovici}.  All this evidence
suggests that criticality -- with its concomitant power laws and scale
invariance -- might play a relevant role in intact-brain dynamics
\cite{Beggs08,Chialvo10}.  At variance with inanimate matter -- for
which the emergence of generic or self-organized criticality in
sandpile models, type-II superconductors, or solar flares is
relatively well understood, \cite{Bak,SOC-Jensen,BJP} -- criticality
in living systems can be conjectured to be the result of evolutionary
or adaptive processes, which for reasons to be understood select for
it.

The criticality hypothesis \cite{Beggs08,Chialvo10,Mora-Bialek} states
that biological systems can perform the complex computations that they
require to survive only by operating at criticality (the edge of
chaos), i.e. at the borderline between an active or chaotic phase in
which noise propagates unboundedly---thereby corrupting all
information processing or storage---and a quiescent or ordered phase
in which perturbations readily fade away, hindering the ability to
react and adapt \cite{Langton,Ber-Nat}. Critical dynamics provides a
delicate trade-off between these two impractical tendencies, and it has
been argued to imply optimal transmission and storage of information
\cite{Haldeman05,Plenz07,Beggs08}, optimal computational capabilities
\cite{Legenstein07}, large network stability \cite{Ber-Nat}, maximal
variety of memory repertoires \cite{BP04}, and maximal sensitivity to
stimuli \cite{Kinouchi-Copelli}. 

Such a delicate balance occurs just at a singular or critical point,
requiring a precise fine tuning.  However, a very recent fMRI analysis
of the human brain at its resting state, reveals that the brain spends
most of the time wandering around a broad region near a critical
point, rather than just sitting at it \cite{Taglia}.  This suggests
that the region where cortical networks operate is not just a critical
point, but a whole extended region around it. 

Here, inspired by this empirical observation as well as by some recent
findings in network theory and neuroscience \cite{GPCN,Rubinov,Zhou12}
we scrutinize the dynamics of simple models of neural activity
propagation when the structural architecture of brain networks is
explicitly taken into account.  Using a combination of analytical and
computational tools, we show that the intrinsically disordered
(hierarchical and modular) organization of brain networks dramatically
influences the dynamics by inducing the emergence -- in the jargon of
Statistical Mechanics -- of a Griffiths phase (GP)
\cite{Griffiths,Noest,Vojta-Review,GPCN}.  This phase, which stems
from the presence of disorder (structural heterogeneity here) is
characterized by generic power-laws extending over broad
regions in parameter space. Furthermore, functional advantages usually ascribed to
criticality, such as a huge sensitivity to stimuli, are reported to
emerge generically all along the GP.  Remarkably, not only do we find
GPs in stylized models of brain architecture, but also in real neural
networks such as those of the {\it C. elegans} and the human
connectome.

Our conclusion is that, as a consequence of the
intrinsically-disordered architecture of brain networks, critical-like
regions are extended from a singular point to a broad or stretched
region, much as evidenced in recent fMRI experiments.  The existence
of Griffiths phases facilitates the task of self-organizing, adaptive,
or evolutive mechanisms seeking for critical-like attributes, with all
their alleged functional advantages.

We claim that the intrinsic structural heterogeneity of cortical
networks calls for a change of paradigm from the
critical/edge-of-chaos to a new one, relying on extended critical-like
Griffiths regions.  Our work also raises a series of questions worthy
of future pursuit. For instance, is there any connection between GPs
and the empirically reported generic power-law decay of short-time
memories?  Do our results extend to other hierarchical architectures
such as those encountered in metabolic or technological networks?


\begin{figure*}[h]
\begin{center}
\vspace{-0cm}
   \includegraphics*[width=16cm]{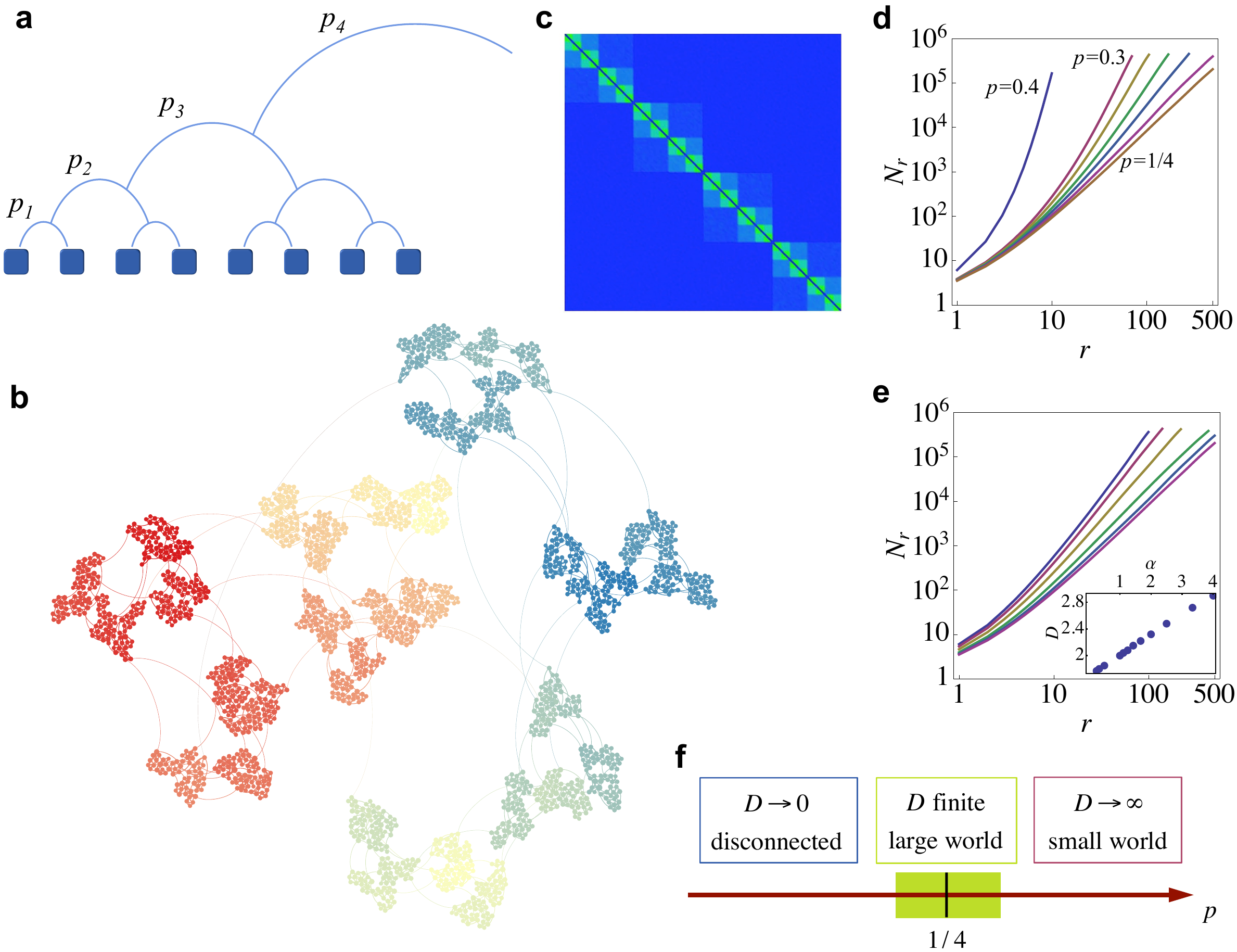}
   \caption{ {\bf Hierarchical modular networks.} ({\bf a}) Sketch of the
     bottom-top approach (HMN-1 model): initially, nodes are grouped
     into fully connected modules of size ${\it M}_\textrm{0}$
     (blue squares); then nodes in different modules are clustered
     recursively into sets of ${\it b}$ higher level blocks (e.g. in
     pairs, ${\it b}=2$) linking their respective nodes with
     hierarchical-level dependent wiring probabilities (HMN-1): ${\it
       p}_\textrm{l} = {\it \alpha} {\it p}^l$ with $0< {\it p} <1$
     and ${\it \alpha}$ a constant.  At level ${\it l}$, each of the
     existing ${\it N}/2^l$ pairs is connected on average by ${\it
       n}_\textrm{l}= 4^l ({\it M}_\textrm{0}/2)^2 \times {\it\alpha
       p}^l $ links.  The resulting networks are always connected;
     with total number of ${\it N} ={\it M}_\textrm{0} \times {\it
       b}^{s}$ nodes, and average connectivity $ {\it k} = ({\it
       M}_\textrm{0}-1)+{\it\alpha} ({\it M}_\textrm{0}/2)\sum_{i=1}^s
     (2 {\it p})^i$. ({\bf b}) Graph representation of a HMN-1 with ${\it
       N}=2^{11}$ nodes, organized across ${\it s}=10$ hierarchical
     levels (${\it M}_\textrm{0}=2$, ${\it p}=1/4$, and ${\it
       \alpha}=4$).  ({\bf c}) Adjacency matrix of the connection-density
     (as in {\bf b}) averaged over several network realizations (greener for
     larger densities).  ({\bf d} and {\bf e}) Topological dimension, $ {\it D}$,
     as defined by ${\it N}_{\it r} \sim {\it r}^D$ (see main text)
     as a function of parameters. ({\bf d}) As ${\it p}$ is increased ${\it
       D}$ (the slope of the straight lines in the double logarithmic
     plot) grows and eventually becomes infinite; for smaller values
     of ${\it p}$ (not shown) it becomes flat, and ${\it D}
     \rightarrow 0$. ({\bf e}) keeping ${\it p} = 1/4$, the topological
     dimension ${\it D}$ is finite and continuously varying as
     function of ${\it \alpha}$, (values summarized in the inset). ({\bf f})
     Summary of structural properties: networks are disconnected
     (vanishing topological dimension, ${\it D}$) and small-world
     (${\it D}=\infty$) for small and large values of ${\it p}$
     respectively, while around ${\it p} = 1/4$ networks have a finite
     ${\it D}$ (as well as a finite connectivity and a finite density
     of connections in the large-${\it N}$ limit, i.e.  networks are
     scalable). }
    \label{fig:Topology}
\end{center}
\end{figure*}

\section{Results}
\subsection{Hierarchical network architectures}
\label{section:HMN}

The cortex network has been the focus of attention in neuroanatomy for
a long time, but only recently, the development of high-throughput
methods has allowed the unveiling of its intricate architecture or
connectome \cite{Hagmann,Honey09}. Brain networks have been found to
be structured in moduli --each modulus being characterized by having a
much denser connectivity within it than with elements in other
moduli-- organized in a hierarchical fashion across many scales
\cite{Sporns,Review-Kaiser,Review-Bullmore}. Moduli exist at each
hierarchical level: cortical columns arise at the lowest level,
cortical areas at intermediate ones, and brain regions emerge
at the systems level, forming a sort of fractal-like nested structure
\cite{Sporns,Review-Kaiser,Review-Bullmore,SCKH}.

To be able to perform systematic analyses, we have designed synthetic
hierarchical and modular networks (HMN) with ${\it s}$ hierarchical
levels, ${\it N}$ nodes/neurons, and ${\it L}$ links/synapses, whose
structure can be tuned to mimic that of real networks.  We employ two
different HMN models based on a bottom-top approach; in the first,
local fully-connected moduli are constructed and then recursively
grouped by establishing new inter-moduli links between them in either
a stochastic way with a level-dependent probability ${\it p}$ as
sketched in Figure 1 (HMN-1) section or in a deterministic way with a
level-dependent number of connections (HMN-2). For further details see
the Materials and Methods (MM) section.  Similarly, top-down
models can also be designed \cite{Zhou11}.

A way to encode key network structural information is the topological
dimension, $ {\it D}$, which measures how the number of neighbors of
any given node grows when moving $1$, $2$, $3$, ..., $ {\it r}$ steps
away from it: $ {\it N_r} \sim {\it r}^D$ for large values of ${\it
  r}$.  Networks with the small-world property \cite{Barabasi-Review}
have local neighborhoods quickly covering the whole network, i.e.
${\it N_r}$ grows exponentially with $ {\it r}$, formally
corresponding to $ {\it D} \rightarrow \infty$.  Instead, large-worlds
have a finite topological dimension, while $ {\it D} =0$ describes
fragmented networks (see Figure 1). Our synthetic HMN models span all
the spectrum of $ {\it D}$-values as illustrated in
Figure 1.
 
Strictly speaking, the HMN networks that we will consider in the following
are finite dimensional only for ${\it p}=1/4$, in which case
the number of inter-moduli connection is stable across hierarchical
levels (see MM). For ${\it p}>1/4$ (resp. ${\it p}<1/4$) networks become more and
more densely (resp. sparsely) connected as the hierarchy depth
(i.e. the network size) is increased.  Deviations from ${\it p}=1/4$
create fractal-like networks up to certain scale, being good
approximations for finite-dimensional networks in finite size.  In
some works (e.g. \cite{PNAS-Gallos}), the Hausdorff (fractal)
dimension ${\it D}_\textrm{f}$ is computed for complex networks.  We have
verified numerically that ${\it D}_\textrm{f}\approx {\it D}$ in all cases
for HMNs.

 \begin{figure*}[t!]
\begin{center}
\vspace{-0cm}
     \includegraphics*[width=16cm]{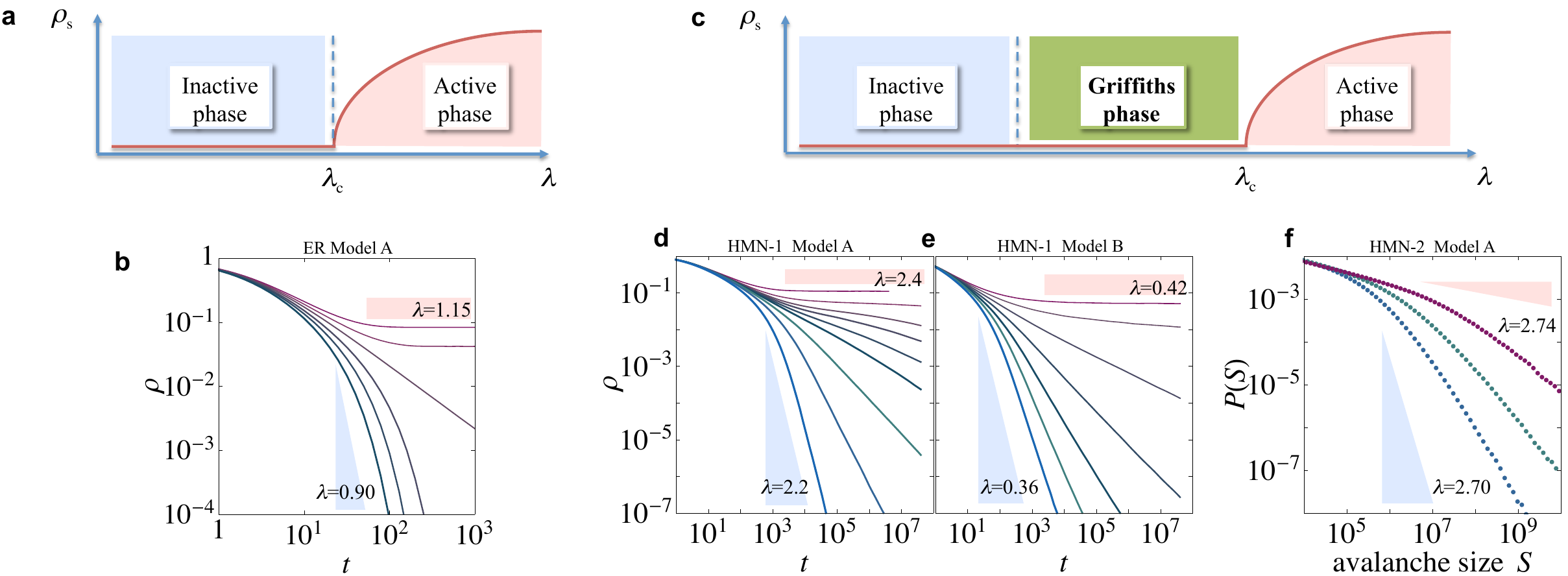}
     \caption{
     {\bf Conventional versus non-conventional phase diagram.}
       Left: ({\bf a} and {\bf b}) conventional critical point scenario for our
       Model A running upon a random Erd\H{o}s-R\'enyi network ($10^6$
       nodes, average connectivity $20$, infinite topological
       dimension $ {\it D}$): a singular power-law separates a quiescent
       phase (with exponential decay) from an active one (with a
       non-trivial steady state). Right: ({\bf c}--{\bf f}) Emergence of broad regions of
       power-law scaling in HMNs. ({\bf c}) Schematic phase diagram for a
       system exhibiting a broad region of power-law scaling.  The
       stationary density of activity, ${\it \rho}_\textrm{s}$, is depicted as a
       function of the spreading rate ${\it \lambda}$.  ({\bf d} and {\bf e}) Steady
       state density of active sites, for Model-A  and 
       Model-B dynamics respectively, on HMN-1 networks (with ${\it N}=2^{14}$ nodes,
       and parameters ${\it s}=13$, ${\it p} =1/4$, ${\it \alpha}=1$). Data for
       increasing values of the spreading rate ${\it \lambda}$, from
       bottom to top. ({\bf f}) Avalanche-size distributions for Model A on
       a HMN-2 network (${\it N}=2^{14}$, ${\it s}=13$, ${\it p
       }=1/4$, ${\it \alpha}=1$;
       Griffiths phase for such networks is observed for
       $2.60\le {\it \lambda} \le2.79$); avalanche sizes are power-law
       distributed over a wide range ${\it \lambda}$ values reflecting the
       existence of a GP.  These conclusions have been confirmed in
       finite-size scaling analyses, and can be generalized for other
       combinations of network architectures and dynamical models.
     }
    \label{fig:Dynamics}
\end{center}
\end{figure*}
 
\subsection{Architecture-induced Griffiths phases}
Disorder is well-known to radically affect the behavior of phase
transitions (see \cite{Vojta-Review} and refs. therein).  In
disordered systems, there exist local regions characterized by
parameter values which differ significantly from their corresponding
system averages. Such rare regions can, for instance, induce the
system to be locally ordered, even if globally it is in the disordered
phase. In this way, in propagation-dynamic models, activity can
transitorily linger for long times within rare active regions, even if
the system is in its quiescent phase. In the particular case in which
broadly different rare-regions exist -- with broadly distinct sizes
and time-scales -- the overall system behavior, determined by the
convolution of their corresponding highly-heterogeneous contributions,
becomes anomalously slow (see Section \ref{section:dynamics}). 
In contrast with standard critical
points, systems with rare-region effects have an intermediate broad
phase separating order from disorder: a Griffiths phase (GP)
with generic power-law behavior and other anomalous properties
(e.g. \cite{Vojta-Review} and Section \ref{section:dynamics})

Remarkably, it has been very recently shown that structural
heterogeneity can play in networked systems a role analogous to that
of standard quenched disorder in physical systems \cite{GPCN}. In
particular, simple dynamical models of activity propagation exhibit
GPs when running upon networks with a finite topological dimension
${\it D}$. On the other hand, in small-world networks (with $ {\it
  D}=\infty$) local neighborhoods are too large -- quickly covering
the whole network -- as to be compatible with the very concept of rare
(isolated) regions \cite{GPCN}. Therefore, it has been conjectured
that a finite topological dimension ${\it D}$ is an excellent
indicator of eventual rare-region effects and GPs.

\subsection{Anomalous propagation dynamics in HMNs} 
\label{section:dynamics}
To model the propagation of neuronal activity, we consider
highly-simplified dynamical models running upon HMNs.  More realistic
models of neural dynamics with additional relevant layers of
information could be considered; but we do not expect them to
significantly affect our conclusions. Our approach here consists
in modeling activity propagation in a minimal way; 
in some of the cases that we study, the network nodes are
not neurons but coarse neural regions, for which effective models of
activity propagation are expected to provide a sound description 
of large-scale properties. 
Every node (or neuron) is endowed with a binary state variable 
${\it \sigma}$, representing either activity ($ {\it \sigma} =1$) or
quiescence (${\it \sigma}=0$).  Each active neuron is spontaneously
deactivated at some rate ${\it \mu}$ (${\it \mu}=1$ here), while it
propagates its activity to other directly connected neurons at rate
${\it \lambda}$. We have considered two different dynamics: in the
first one, ({\bf Model A}) a synapsis between an active and a quiescent
node is randomly selected at each time and proved for activation,
while in the other variant ({\bf Model B}) a neuron is selected
and all its neighbors are proved for activation. Details of the
computational implementation of the two models, known in statistical
physics as the contact process and the SIS model respectively, can be
found in MM.

In general, depending on the value of $\lambda$ these models can be
either in an active phase --for which the density ${\it \rho}$ of
active nodes reaches a steady-state value ${\it \rho}_\textrm{s} > 0$
in the large system-size and large-time limit-- or in the inactive
phase in which ${\it \rho}$ falls ineluctably in the quiescent
configuration (${\it\rho}_\textrm{s} = 0$). Separating these two
regimes, at some value, ${\it\lambda}_\textrm{c}$, there is a standard
critical point where the system exhibits power-law behavior for
quantities of interest, such as the time decay of a homogeneous
initial activity density, ${\it\rho(t)} \sim {\it t}^{-\theta}$, or
the size-distribution of avalanches triggered by an initially localized
perturbation, $P({\it S}) \sim {\it S}^{-\tau}$.  Here ${\it \theta}$ and
${\it \tau}$ are critical indices or exponents. This standard
critical-point scenario (see Figures 2a and 2b) holds for regular
lattices, Erd\H{o}s-Renyi networks, and many other types of
networks. On the other hand, computer simulations of the different
dynamical models running upon our complex HMN topologies with finite
${\it D}$ reveal a radically different behavior (see Figure
2c--2f and MM). The power-law decay of the average density
${\it \rho(t)}$ -- specific to the critical point in pure systems --
extends to a broad range of ${\it \lambda}$ values.  The existence of
a broad interval with power-law decaying activity is supported by
finite size scaling analyses reported in MM. Likewise, as shown in
Figure 2f, avalanches of activity generated from a
localized seed have power-law distributed sizes, with continuously
varying exponents, in the same broad region. These features are
fingerprints of a GP and have been confirmed to be robust against
increasing system size (up to ${\it N}=2^{20}$), using different types
of HMN (HMN-1 with different values of ${\it \alpha}$ and ${\it p}$,
HMN-2 models, all with finite ${\it D}$) and dynamical models (see MM).

How do Griffiths phases work?
For illustration purposes, let us consider a simplified example. 
Consider Model A (the contact process) on a generic network, with a
node-dependent quenched spreading rate ${\it \lambda}({\bf x})$,
characterized --without loss of generality-- by a bimodal distribution
of $\lambda$ with average value $\bar{{\it \lambda}}$. Suppose the two
possible values of $\lambda$ are one above and one below the critical
point of the pure model, ${\it \lambda}_{\textrm c} $. In this way, at
each location the system has an intrinsic preference to be either in
the active or in the quiescent phase. Under these circumstances,
typically, ${\it \lambda}_\textrm{c} < \bar{{\it \lambda}}_\textrm{c}
$, so that, for values of $\bar{{\it\lambda}}$ in between ${\it
  \lambda}_\textrm{c}$ and $\bar{{\it \lambda}}_\textrm{c}$ the
disordered system is in its quiescent phase. However, there are always
spatial locations characterized by significantly over-average values
of (actually, local values of ${\it\lambda}({\bf x}) > {\it
  \lambda}_\textrm{c} $).  In these regions, initial activity can
linger for very long periods, especially if they happen to be large.
Still, as such rare regions have a finite size, they ineluctably
end up falling into the inactive state.  Considering a rare active
region of size ${\it \zeta}$, it decays to the quiescent state after a
typical time ${\it \tau(\zeta)}$, which grows exponentially with cluster
size, i.e.  ${\it \tau}\simeq t_\textrm{0} \exp[{\it A(\lambda) \zeta]}$
(Arrhenius law), where ${\it t}_\textrm{0}$ and ${\it A(\lambda)}$ do
not depend on ${\it \zeta}$.  On the other hand, the distribution of ${\it
  \zeta}$-values is also exponential (very large regions are exponentially
rare).  Therefore, the overall activity density, ${\it \rho(t)}$,
decays as the following convolution integral 
\begin{equation}
{\it\rho(t)} \sim \int
d{\it \zeta} P({\it \zeta}) {\it \zeta}\exp{[-{\it t}/({\it t}_\textrm{0}
  e^{A(\lambda) \zeta})]},
\end{equation} which evaluated in saddle-point approximation
leads to ${\it \rho(t)} \sim {\it t}^{-\theta}$, with ${\it
  \theta}(\bar{ {\it \lambda)}}$
varying continuously with the disorder average value, $\bar{{\it
    \lambda}}$.  Such generic power laws signal the emergence of
Griffiths phases.  This is just an explanatory example of a general phenomenon,
thoroughly studied in classical, quantum, and non-equilibrium
disordered systems \cite{Vojta-Review}. In HMNs, the quenched 
disorder is encoded in the intrinsic disorder of the hierarchical 
contact pattern.

\begin{figure}
\begin{center}
\includegraphics*[width=8cm]{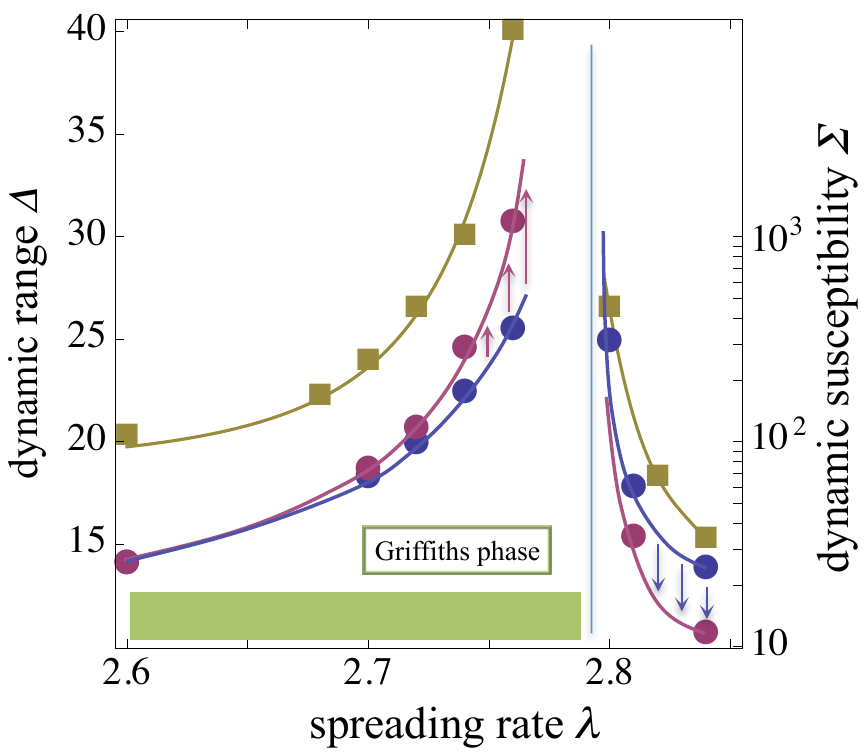}
\caption{
{\bf Network response diverges all along the Griffiths
    phase.} Right axis: dynamic susceptibility, ${\it \Sigma}$,
  measured for the dynamical Model A in HMN-2 networks of size ${\it
    N}=2^{14}$ (blue circles) and ${\it N}=2^{17}$ (purple circles);
  the critical point ${\it \lambda}_\textrm{c} \approx 2.79$, is
  marked as a vertical line (see MM: dynamical protocol iii). In the
  GP (resp. active phase) the overall response increases (resp. decreases)
  for larger systems as marked by the red (resp. blue) arrows.  In the
  GP, ${\it \Sigma}$ grows as a ${\it \lambda}$-dependent power-law of
  ${\it N}$.  Left axes: dynamic range ${\it \Delta}$ (green squares)
  for the same case as above, ${\it N}=2^{14}$ (dynamical protocol iv;
  see MM).  Instead of the usual symmetric cusp-singularity at the
  critical point, there is a whole region of extremely large ${\it
    \Delta}$ with a strong asymmetry around the transition point, as
  corresponds to the existence of a GP.
}
\label{fig:Response}
\end{center}
\end{figure}

\subsection{Diverging response in HMNs} One of the main alleged
advantages of operating at criticality is the strong enhancement of
the system's ability to distinctly react to highly diverse stimuli. In
the statistical mechanics jargon this stems from the divergence of the
susceptibility at criticality \cite{Binney}.  How do systems with
broad GPs respond to stimuli?  To answer this question we measure the
following two different quantities (see MM):

{\bf(i)} The dynamic susceptibility gauges the overall response
to a continuous localized stimulus and is defined as
${\it\Sigma(\lambda)}= {\it N}[{\it\rho}_\textrm{f}({\it \lambda})-
{\it\rho}_\textrm{s}({\it\lambda})]$, where
${\it\rho}_\textrm{s}({\it\lambda})$ is the stationary density in the
absence of stimuli and ${\it\rho}_\textrm{f}({\it\lambda})$ is the
steady-state density reached when one single node is constrained to
remain active.  As shown in Figure 3, ${\it\Sigma}$
becomes extremely large in the GP and, more importantly, it grows as a
power-law of system-size, ${\it\Sigma}({\it\lambda}\approx
{\it\lambda}_\textrm{c},{\it N})\sim {\it N}^\eta$, implying that
there is an extended extended region (the whole GP) where the system
exhibits a divergent response (with ${\it \lambda}$-dependent
continuously varying exponents)

{\bf(ii)} The dynamic range, ${\it \Delta}$, introduced in this
context in \cite{Kinouchi-Copelli}, measures the range of perturbation
intensities/frequencies for which the system reacts in distinct ways,
being thus able to discriminate among them.  We have computed ${\it
  \Delta}({\it \lambda})$ in the HMN-2 model (see Figure
3) which clearly illustrates the presence of a broad
region with huge dynamic ranges (rather than the standard situation
for which large responses are sharply peaked at criticality).

Therefore, if critical-like dynamics is important to access a broad
dynamic range and to enhance system's sensitivity, then it becomes
much more convenient to operate with hierarchical-modular systems,
where criticality -- with extremely large responses and huge
sensitivity -- becomes a region rather than a singular point.

\begin{figure}
\begin{center}
     \includegraphics*[width=7.5cm]{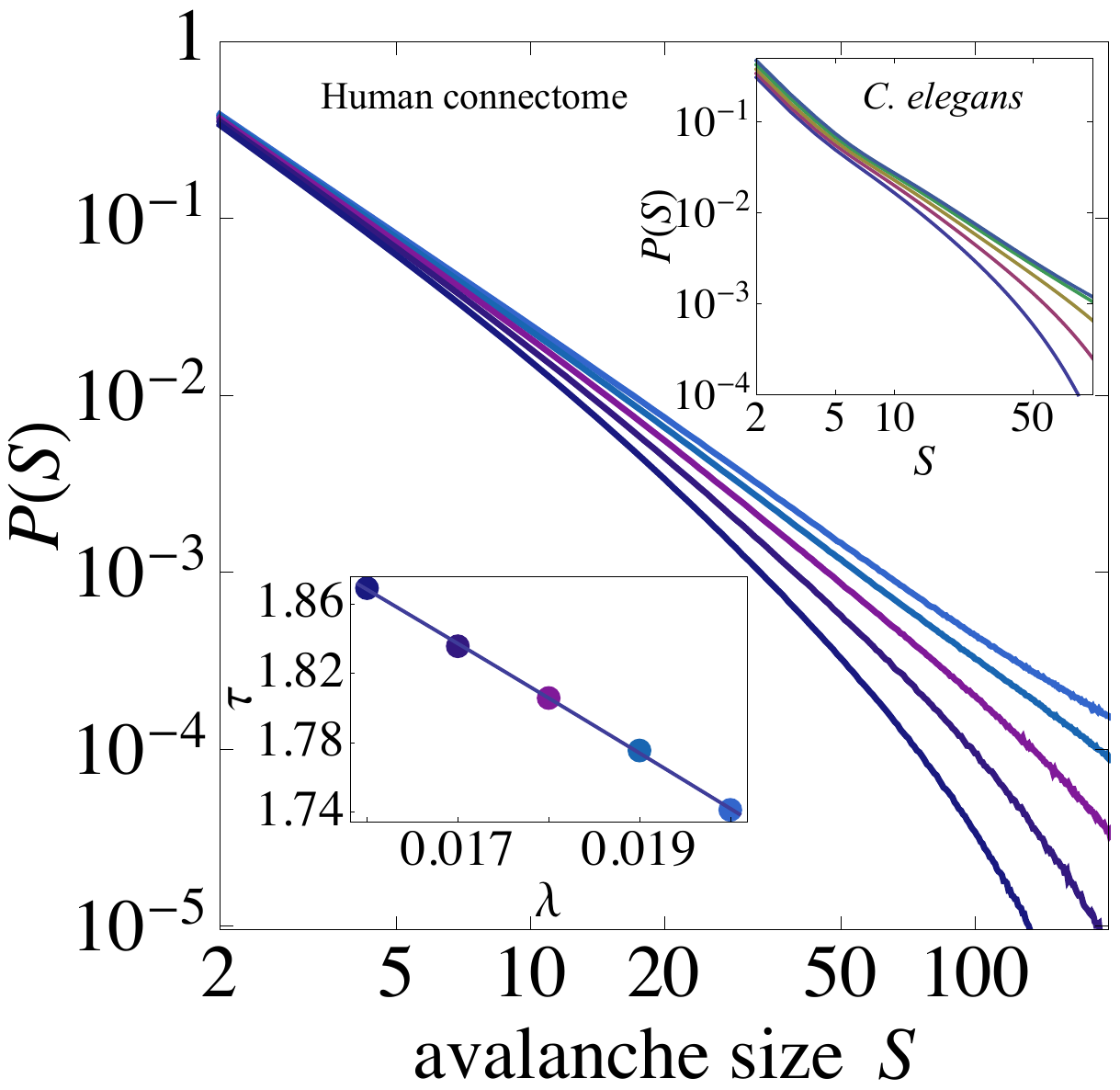}
     \caption{
     {\bf Critical avalanches all along the Griffiths phase
         in real brain networks.} Avalanche-size statistics for Model-B
       dynamics in real neural networks. Main plot: human connectome
       network (consisting of $998$ brain areas and the structural
       connections among them). Avalanche-size distributions for
       Model-B dynamics (with $0.016 \leq \lambda \leq 0.020$
       equispaced values).  Truncated power-laws $P({\it S})\sim {\it
         S}^{-\tau}e^{-S/\xi}$ with continuously varying exponents
       ${\it \tau}$ constitute the most reliable fits according to the
       Kolmogorov-Smirnoff criterion (see MM). Lower inset: The ${\it
         \tau}$ exponent (estimated through non-linear least-square
       fits) as a function of ${\it \lambda}$; it converges to
       $\approx 1.7$ at the critical point, ${\it
         \lambda}_\textrm{c}\approx 0.020$. Upper inset: as the main
       figure but for the {\it C. elegans} neural network ($0.08 \leq
       {\it \lambda} \leq 0.12$).
      }
    \label{fig:RealNetworks}
\end{center}
\end{figure}

\subsection{Griffiths phases in real networks}         
Analyses of different nature have revealed that organisms from the
primitive {\it C. elegans}
\cite{Review-Bullmore,Review-Kaiser,Sinha,Varshney} (for which a full
detailed map of its about $300$ neurons has been constructed) to cats,
macaque monkeys, or humans (for which large-scale connectivity maps
are known \cite{Hagmann}) have a hierarchically organized neural
network. Such structure is also shared by functional brain networks
(e.g. from functional magnetic resonance imaging data)
\cite{Zhou06,PNAS-Gallos,Hagmann,Honey09,Bassett10}.  Do simple
dynamical models of activity propagation (such as those in previous
sections) running upon real neural networks exhibit GPs?

We have considered the human connectome network, obtained by Sporns
and collaborators using diffusion imaging techniques
\cite{Hagmann,Honey09}. It consists of a highly coarse-grained mapping
(as opposed for instance to the detailed map of {\it C. elegans}) of
anatomical connections in the human brain, comprising ${\it N}=998$
brain areas and the fiber tract densities between them, with a
hierarchical organization \cite{Review-Bullmore,Review-Kaiser} (see
MM).

Given that this network comprises only ${\it N}\lesssim 1000$ nodes,
the maximum size of possible rare regions and the associated
power laws are necessarily cut-off at small sizes and short
times. Nevertheless, as illustrated in
Figure 4, simulations of the dynamical models
above (Model B in this case) show a significant deviation from the typical standard
critical-point scenario. Instead, avalanches are clearly distributed
as power laws, with moderate finite-size effects, in a broad range of
${\it \lambda}$-values (see Figure 4).  Actually,
truncated power-laws of the form $P({\it S})\sim {\it S}^{-\tau}e^{-
  S/\xi}$ -- with ${\it \lambda}$-dependent values of ${\it \tau}$ --
provide highly reliable fits of the size distributions, $P({\it S})$,
according to the Kolmogorov-Smirnoff criterion (see MM), supporting
the picture of a broad critical-like region. This strongly suggests
that if it were feasible to run dynamical models upon the actual human
brain network (with about $10^{12}$ neurons and $10^{15}$ synapses) a
GP would appear in a robust way, and would extend over much larger
range of size and time scales. Similar results, even if affected by
more severe size effects, are obtained for the {\it C. elegans}
detailed neural network, consisting of ${\it N}\lesssim 300$ neurons
(see Figure 4, upper inset).

\subsection{Spectral fingerprints of Griffiths phases in HMNs}

To further confirm the existence of Griffiths phases (beyond direct
computational simulations), here we present some analytical results.
An important tool in the analysis of network dynamics is provided by
spectral graph theory, in which the network structure is encoded in
some matrix and linear algebra techniques are exploited
\cite{FanChung}.  For instance, the dynamics of simple 
models (e.g. Model-B) is often governed by the largest (or principal)
eigenvalue, ${\it \Lambda}_\textrm{max}$, of the adjacency matrix,
${\it A}_{ij}$ (with $1$'s as entries for existing connections
and $0$'s elsewhere), which straightforwardly appears in a standard
linear stability analysis (as detailed in the MM section).  It is easy
to show that (with very mild assumptions) the critical point
--signaling the limit of linear stability of perturbations on the
quiescent state-- is given by ${\it \lambda}_\textrm{c}
{\it\Lambda}_\textrm{max}=1$. Remarkably, it has been recently shown that this general result may
not hold for certain networks, for which the largest eigenvalue has an
associated localized eigenvector (i.e. with only a few non-vanishing
components; this phenomenon closely resembles Anderson localization in
physics \cite{Goltsev12}). In such a case, linear instabilities for
${\it \lambda}$ slightly larger than $1/{\it \Lambda}_\textrm{max}$
lead to localized activity around only few nodes (where the
corresponding eigenvector is localized), not pervading the network and
not leading to a true active state. This implies that the critical
point is shifted to a larger value of ${\it \lambda}$. Instead, these
regions of localized activity resemble very much rare-regions in GPs:
activity lingers around them until eventually a large fluctuation
kills them.

\begin{figure}
\begin{center}
     \includegraphics*[width=15cm]{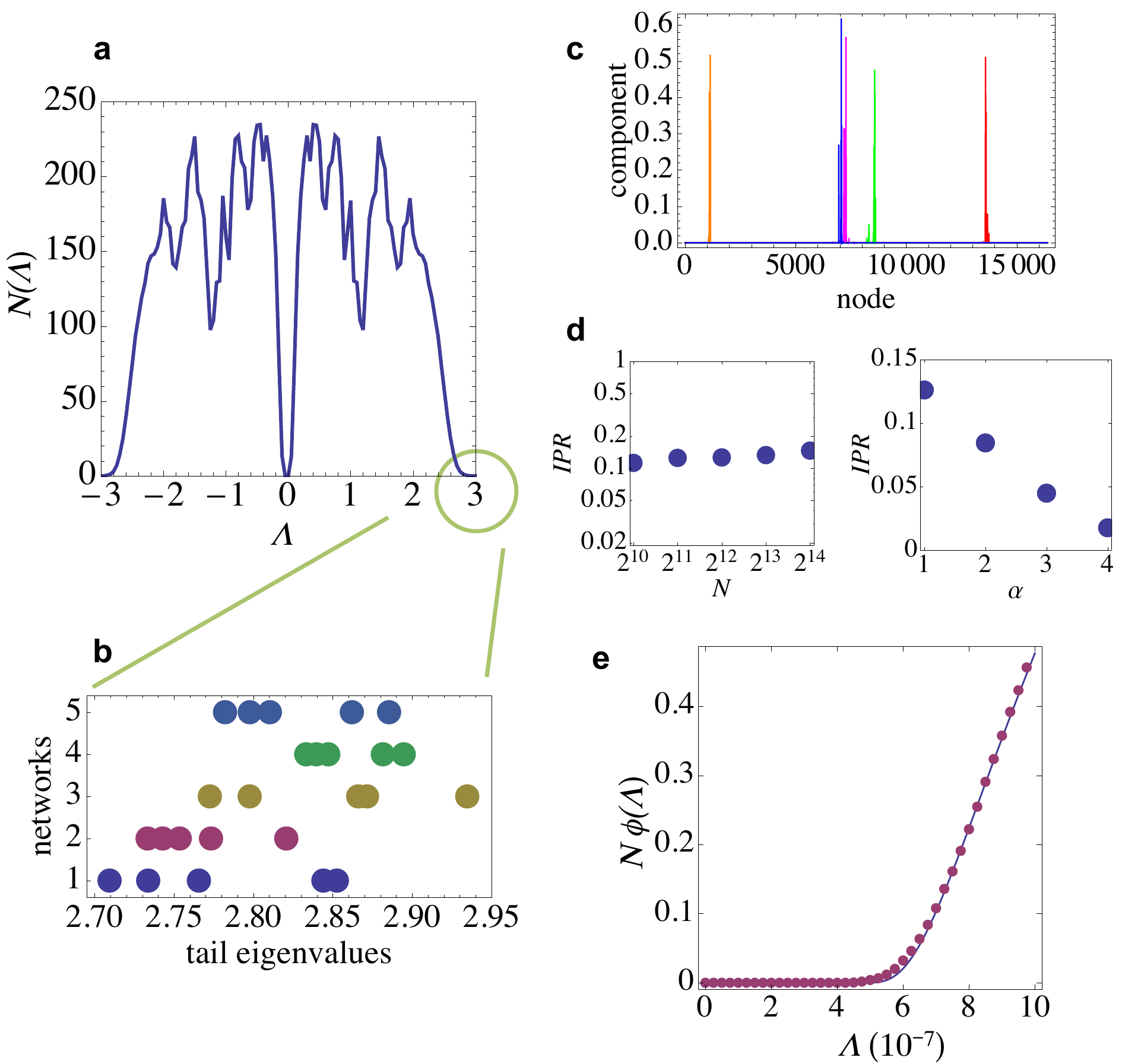}
     \caption{
     {\bf Spectral analyses of hierarchical networks.}  ({\bf a})
    Average spectrum of the adjacency matrix of HMN-2 networks (${\it
      N}=2^{14}$, ${\it s}=13$, $ {\it M}_\textrm{0}=2$, ${\it
      \alpha}=1$). Data are averaged over $150$ network
    realizations. The vertical axis reports the average number of
    eigenvalues (not the density). ({\bf b}) Ideal zoom of the higher
    spectral edge of ({\bf a}). The values of the $5$ largest eigenvalues
    for $5$ randomly chosen networks from ({\bf a}) are represented. No
    proper spectral gap is observed. ({\bf c}) Localization of the $5$
    eigenvectors corresponding to the largest eigenvalues. The
    principal eigenvector ${\bf f}({\it \Lambda}_\textrm{max})$ is plotted in
    red. Being our HMNs connected, in agreement with the
    Perron-Frobenius theorem \cite{gantmacher} the components of ${\bf
      f}({\it \Lambda}_\textrm{max})$ (even the vanishing ones) are all
    strictly positive, although this cannot be appreciated in linear
    scale. The next eigenvectors are plotted in magenta, orange, green
    and blue.  ({\bf d}) Dependence of the IPR on the system size and the
    number of block-block connections in HMN-2 networks. ({\bf e}) Lower
    spectral edge of the cumulative distribution of Laplacian
    eigenvalues of HMN-2 networks as in ({\bf a}). Numerical data (points)
    are compared to an exponential Lifshitz tail with exponent ${\it
      a}\approx 1.00$. The Laplacian matrix is defined as ${\it
      L}_{ij}=\sum_{k}A_{ik}-A_{ij}$. }
\end{center}
\end{figure}

Inspired by this novel idea, we performed a spectral analysis of our
HMNs (e.g. HMN-2 nets with ${\it \alpha}= 1$, see Figure 5), with the result that --
for finite ${\it D}$ networks -- not only the largest eigenvalue ${\it
  \Lambda}_\textrm{max}$ corresponds to a localized eigenvector, but a whole
range of eigenvalues below ${\it \Lambda}_\textrm{max}$ (even hundreds of
them) share this feature, as can be quantitatively confirmed (see Figure 5c-d and
MM). In particular, the principal eigenvector is heavily peaked around
a cluster of neighboring nodes.  We have conjectured and verified
numerically that the clusters where the largest eigenvalues are
localized correspond to the rare-regions, with above-average
connectivity and where localized activity lingers for long time.
Also, we numerically found ${\it \lambda}_\textrm{c} \approx 0.41 >
1/\Lambda_\textrm{max} \approx 0.33 $, confirming the prediction above. The
interval between these two values defines the GP (see MM).

In addition, we also considered large network ensembles and computed the
probability distribution of eigenvalues.  
We found that the distribution of the eigenvalues corresponding to 
localized eigenvectors results in an exponential tail of the continuum spectrum,
where an infinite dimensional graph would exhibit a spectral gap 
instead (see MM).  This translates into an exponential tail of the cumulative
distribution $\phi({\it\Lambda})$ of Laplacian eigenvalues (or
  integrated density of states) at the lower spectral edge, a
so-called Lifshitz tail -- which in equilibrium systems is
related to the Griffiths singularity \cite{Nie}. We have found
Lifshitz tails with their characteristic form
\begin{equation}
\phi({\it \Lambda}) \propto \exp \left[ -\frac{1}{({\it \Lambda}-{\it\Lambda_\textrm{min}})^a}\right]
\end{equation}
(where $a$ is a real parameter, see Figure 5e). Interestingly, Lifshitz tails are
also rigorously predicted on Erd\H{o}s R\'enyi networks below the
percolation threshold, where rare-region effects and Griffiths phases
are an obvious consequence of the network disconnectedness
\cite{Khorunzhiy06lifshitstails}.
Therefore, the presence
of both (i) localized eigenvectors and (ii) Lifshitz tails confirms
the existence of GPs in networks with complex heterogeneous
architectures.

\begin{figure}
\begin{center}
     \includegraphics*[width=7.5cm]{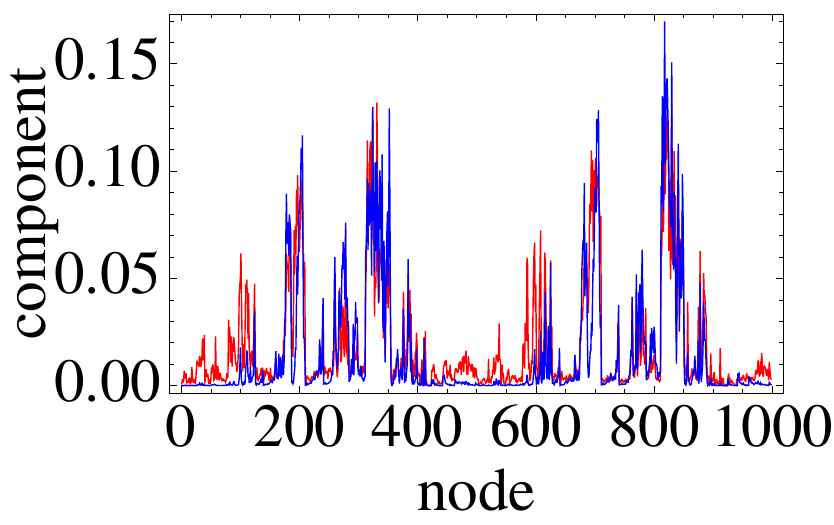}
     \caption{
     {\bf Localization properties of the human connectome
    network.} Principal eigenvector of the weighted (blue) and
    unweighted (red) adjacency matrix. As the peak structure is
    roughly preserved, localization properties are expected to be
    alike in the two representations. In both cases regions of
    localized activity are evident.}
\end{center}
\end{figure}

The fingerprints of extended criticality in the human connectome are a
result of its hierarchical network structure, and the localization
properties that characterize it. Figure
6 supports this view, highlighting the
localization of the principal eigenvector. In particular, we show that
the principal eigenvector of the full adjacency matrix and that of the
unweighted (i.e.  with $1$'s as entries in correspondence to every
non-zero weight) version of it display very similar peak structures
(i.e. their rare regions are very similar). This last observation
supports our choice to run simple unweighted dynamics on top of the
connectome network. Connection weights will certainly contribute to a
fully realistic description of the brain function, but they are not
necessary in order to achieve broad criticality, which primarily
arises from structural disorder.

 \section{Discussion}
 A few pioneering studies have recently explored the role of
 hierarchical modular networks (HMN) on different aspects of neural
 dynamics.  Rubinov {\it et al.} \cite{Rubinov} argued that HMNs play
 a crucial role in fostering the existence of critical dynamics.
 Similar observations have been also made for spreading dynamics
 \cite{Kaiser07,Muller08,Zhou11} and for self-organization models
 \cite{Zhou12}, but so far these remain empirical findings lacking a
 satisfactory explanation.

 Aimed at shedding light on these issues, here we have made the
 conjecture that owing to the intrinsically heterogeneous
 (i.e. disordered) architecture of brain HMNs and their large-world
 nature (which implies finite topological dimensions ${\it D}$),
 Griffiths phases are expected to emerge. These are
 characterized by broad regions in parameter space exhibiting
 critical-like features and, thus, not requiring of a too-precise fine
 tuning. To confirm this claim, we use a combination of computational
 tools and analytical arguments. In particular, we have constructed
 different synthetic HMNs covering a broad range of architectures
 (with either large- or small-world properties).  For large worlds
 (i.e.  finite topological dimension ${\it D}$), we find ({\bf i})
 generic anomalous slow relaxation and power-law distributed
 avalanches (with continuously varying exponents) in a broad region
 comprised between the active and the inactive phase; ({\bf ii}) the
 system's response to stimuli (as measured by the dynamic
 susceptibility and the dynamic range) is anomalously large: it
 diverges with system size through the whole GP.  At a theoretical
 level, graph-spectral analyses reveal the presence of localized
 eigenvectors and Lifshitz tails which are the spectral counterparts
 of rare regions and GPs. All these evidences confirm the existence of
 GPs in synthetic HMNs, including a spectral graph theory viewpoint,
 thereby going beyond previous results in generic complex networks
 \cite{GPCN}.
 
 More remarkably, we have also provided evidence that GPs appear in
 actual real networks such as those of the human connectome and {\it
   C. elegans}, even if owing to the limited available network-sizes
 the associated power-laws are necessarily cut-off. Still, the best
 fits to data are provided by continuously varying power-laws,
 truncated by finite-size effects.
   
 It is noteworthy that disorder needs to be present at very different
 scales for GPs to emerge (otherwise rare regions cannot be
 arbitrarily large); this is why a hierarchy of levels is required.
 Plain modular networks -- without the broad distribution of cluster
 sizes, characteristic of hierarchical structures -- are not able to
 support GPs.  We emphasize that GPs in HMNs are induced merely by the
 existence of structural disorder: no additional form of neuronal or
 synaptic heterogeneity has been considered here; adding further
 heterogeneity would only enhance rare-region effects and hence GPs.
 For instance, we have also found GPs for the human connectome, when
 taking into account the relative weight of structural connections. It
 should also be noticed that all networks in our study are undirected
 in the sense that links are symmetrical, while synapses in cortical
 networks are not.  Again, this does not jeopardize our conclusions:
 directness in the connections only strengthens isolation and hence
 rare-region effects and GPs.
 
 The existence of a GP with its concomitant generic power-law scaling
 -- not requiring a delicate parameter fine tuning -- provides a more
 robust and solid basis to justify the ubiquitous presence of
 scale-free behavior in neural data, from EEG or MEG records to neural
 avalanches. More in particular, it might give us the key to
 understanding why broad regions around criticality are observed in
 fMRI experiments of the brain resting state \cite{Taglia}.

 As we have shown, the system's response is extremely large (and
 diverges upon increasing the network size) in a broad region of
 parameter space. This strongly facilitates the task of mechanisms
 selecting for the alleged virtues of scale-invariance and strongly
 suggests that a new paradigm is needed: a theory of
 self-organization/evolution/adaptation to the broad region separating
 order from chaos.  In particular, GPs combined with standard
 mechanisms for self-organization \cite{Levina07,Millman,JABO-2} are
 expected to account for the empirically found dispersion around
 criticality \cite{Taglia}.  This also invites the question of whether
 intrinsically disordered HMNs do indeed generically optimize
 transmission and storage of information, improve computational
 capabilities \cite{Legenstein07}, or significantly enhance large
 network stability \cite{Ber-Nat}, without the need to invoke precise
 criticality.

 Usually, brain networks are claimed to be small worlds; instead we
 have shown that large-world architectures are essential for GPs to
 emerge. A solution to this conundrum was provided by Gallos {\it et
   al.} \cite{PNAS-Gallos}, who found that (functional) brain networks
 consist of a myriad of densely connected local moduli, which
 altogether form a large world structure and, therefore, are far from
 being small-world; however, incorporating weak ties into the
 network converts it into a small world preserving an underlying 
 backbone of well-defined moduli. To this end, it is essential that networks
 have a finite Hausdorff or fractal dimension (see Figure 1 and 
 \cite{PNAS-Gallos}), confirming that finite dimensionality is a crucial feature.
 In this way, cortical networks achieve an optimal balance between
 local specialized processing and global integration through their
 hierarchical organization. On the other hand, weak links are not
 expected to significantly affect the existence of GPs. Therefore,
 from this perspective, (i) achieving such an optimal balance and (ii)
 operating with critical-like properties, can be seen as the two sides of
 the same coin. This stresses the need to develop new models for the
 co-evolution of structure and function in neural networks.

It is noteworthy that a mechanism for robust working memory without
synaptic plasticity has been put forward very recently
\cite{Samuel}. It heavily relies on the existence of heterogeneous
local clusters of densely inter-connected neurons, where activity
(memories) reverberates. Not surprisingly, this leads to power-law
distributed fade-away times, which have been claimed to be the
correlate of power-law forgetting \cite{Forgetting}. This is an
eloquent illustration of Griffiths phases at work.

Given that disorder is an intrinsic and unavoidable feature of neural
systems and that neural-network architectures are hierarchical,
Griffiths phases are expected to play a relevant role in many
dynamical aspects and, hence, they should become a relevant concept in
Neuroscience as well as in other fields such as systems biology, where
HMNs play a key role \cite{Escherichia}.  We hope that our work
contributes to this purpose, fostering further research.
\section{Materials and Methods} 

{\bf Synthetic hierarchical networks.}  HMN-1: At each hierarchical
level $ {\it l}=1,2,\ldots s$, pairs of blocks are selected, each
block of size $2^{i-1} {\it M}_\textrm{0}$. All possible $4^{i-1} {\it M}_\textrm{0}^2$
undirected eventual connections between the two blocks are evaluated,
and established with probability ${\it \alpha p}^l$, avoiding
repetitions. With our choice of parameters we stay away from regions
of the parameter space for which the average number of connections
between blocks ${\it n}_l$ is less than one, as this would lead
inevitably to disconnected networks (as rare-region effects would be a
trivial consequence of disconnectedness, we work exclusively on {\it
  connected} networks, i.e. networks with no isolated
components). Since links are established stochastically, there is
always a chance that, after spanning all possible connections between
two blocks, no link is actually assigned. In such a case, the process
is repeated until, at the end of it, at least one link is
established. This procedure enforces the connectedness of the network
and its hierarchical structure, introducing a cutoff for the minimum
number of block-block connections at $1$. Observe also that for ${\it
  M}_\textrm{0}=2$ and ${\it p}=1/4$, ${\it \alpha}$ is the target average
number of block-block connections and $1+{\it\alpha}$ the target
average degree. However, by applying the above procedure to enforce
connectedness, both the number of connections and the degree are
eventually slightly larger than these expected, unconstrained, values.

For the HMN-2, the number of connections between blocks at every level
is {\it a priori} set to a constant value ${\it \alpha}$. Undirected
connections are assigned choosing random pairs of nodes from the two
blocks under examination, avoiding repetitions. Choosing ${\it
  \alpha}\ge 1$ ensures that the network is hierarchical and
connected. This method is also stochastic in assigning connections,
although the number of them (as well as the degree of the network) is
fixed {\it deterministically}.  In both cases, the resulting networks
exhibit a degree distribution characterized by a fast exponential
tail, as shown in Supplementary Figure S1.



 

{\bf Empirical brain networks.} Data of the adjacency matrix of {\it
  C. elegans} are publicly available (see e.g. \cite{Review-Kaiser}).
Different analyses have confirmed that this network has a
hierarchical-modular structure (see
e.g. \cite{Review-Bullmore,Sinha,Varshney,Reese,2Vicsek}).  In
particular, in \cite{2Vicsek} a new measure is defined to quantify the
degree of hierarchy in complex networks. The {\it C. elegans} neural
networks is found to be $6$ times more hierarchical than the average
of similar randomized networks.  Connectivity data for the human
connectome network have been recently obtained experimentally
\cite{Hagmann,Honey09}. In this case too, the network is
hierarchical \cite{Review-Bullmore,Bassett10,Taglia}. Supplementary
Figure S2 shows a graphical representation of its adjacency matrix,
highlighting its hierarchical organization.


 
  
{\bf Dynamical models.}  In both cases (Model A and Model B), neurons
are identified with nodes of the network and are endowed with a binary
state-variable ${\it\sigma}=0,1$. The state of the system is
sequentially updated as follows. A list of active sites is kept.
Model A: At each step, an active node is selected and becomes 
inactive ${\it\sigma}=0$ with probability ${\it\mu/(\lambda+\mu)}$,
while with complementary probability ${\it\lambda/(\lambda+\mu)}$, it
activates one randomly chosen nearest neighbor provided it was
inactive. Model B: At each step, an active node is selected and
becomes inactive ${\it\sigma}=0$ with probability
${\it\mu/(\lambda+\mu)}$, while with complementary probability
${\it\lambda/(\lambda+\mu)}$, it checks all of its nearest neighbors,
activating each of them with probability $0<{\it\lambda}<1$ provided
it was inactive, then it deactivates.  Both Model A and B have
well-known counterparts in computational epidemiology, where they
correspond to the contact process and the
susceptible-infective-susceptible model respectively (see for instance
\cite{Romu01}).  The value of similar minimalistic dynamic rules in
Neuroscience was proven before, e.g. in \cite{Grinstein-Linsker}.
Results for real networks are obtained by running Model B dynamics on
the unweighted version of the network. We have verified that the
introduction of weights does not alter the qualitative picture
obtained.

{\bf Dynamical protocols.} We employ four different dynamical
protocols: {\bf (i)} Decay of a homogeneous state: all nodes are
initially active and the system is let  evolve, monitoring the
density of active sites ${\it \rho (t)}$ as a function of time. {\bf
  (ii)} Spreading from a localized initial seed: an individual node is
activated in an otherwise quiescent network. It produces an avalanche
of activity, lasting until the system eventually falls back to the
quiescent state; the survival probability $P_\textrm{s}({\it t})$ is
measured. The avalanche size ${\it S}$ is defined as the number of
activation events that occur for the duration of the avalanche
itself. The process is iterated and the avalanche size distribution
$P({\it S})$ is monitored.  {\bf (iii)} Identical to (ii), except that
the seed is kept active throughout the simulation (continuous
stimulus). {\bf (iv)} Identical to (ii), except that the seed node is
subsequently reactivated with probability ${\it p}_{\textrm{stimulus}}=1-\exp
(-{\it r} \,{\it \Delta t})$ for ${\it t }>0$ (Poissonian stimulus of
rate ${\it r}$).

{\bf Measures of response.}  The standard method to estimate responses
consisting in measuring the variance of activity in the steady state,
would not provide a measure of susceptibility in the Griffiths region,
where a steady state is trivial (quiescent). We define the dynamic
susceptibility as $ {\it \Sigma}({\it\lambda})={\it
  N}[{\it\rho}_\textrm{f}({\it\lambda})-{\it\rho}_\textrm{s}({\it\lambda})],
$ where ${\it\rho}_\textrm{f}({\it\lambda})$ is the steady state
reached when a single node is constrained to be active throughout the
simulation and ${\it\rho}_\textrm{s}({\it\lambda})$ the steady state
for protocol (i), i.e. in the absence of constraints. In the inactive
state, ${\it\rho}_\textrm{s}({\it\lambda})=0$ while
${\it\rho}_\textrm{f}({\it\lambda})$ is finite but small (of the order
of $1/{\it N}$) as the active node continuously fosters activity in
its surroundings. In the active state,
${\it\rho}_\textrm{s}({\it\lambda})$ and
${\it\rho}_\textrm{f}({\it\lambda})$ are both large and again differ
by a little amount (given by the fixed node and its induced activity)
that vanishes for larger system sizes.  Only in the parameter region
where the response of the system is high, the little perturbation
introduced by the constrained node produces a diverging response.
This is found to occur throughout the Griffiths region (see Figure 3,
main text), confirming the claim of an anomalous response over an
extended range of the parameter ${\it\lambda}$.

An alternative measure of response is provided by the dynamic range
${\it\Delta}$, introduced in Ref. \cite{Kinouchi-Copelli}. We
determine ${\it\Delta}({\it\lambda})$ for various values of ${\it
  \lambda}$ in the Griffiths and active phases, as follows:

(i) a seed node is chosen and initially activated, but not constrained to be
active;
(ii)  the dynamical model (A, B, $\ldots$) is run;
(iii) if the dynamics selects the seed node, and it is found inactive, it is
reactivated with probability ${\it p}_\textrm{stimulus}$;
(iv) the steady-state density ${\it\rho}$ is recorded (due to the intermittent
reactivation, a steady state depending on ${\it p}_\textrm{stimulus}$ is always reached,
unless ${\it p}_\textrm{stimulus}$ is infinitesimal);
(v)  upon varying ${\it p}_\textrm{stimulus}$, the steady-state density ${\it\rho}$ varies
continuously within a finite window. We identify the values ${\it\rho}_{0.1}$ and
${\it\rho}_{0.9}$, corresponding to the $10\%$ and $90\%$ values within such window,
and call ${\it p}_{0.1}$ and ${\it p}_{0.9}$ the values of ${\it p}_\textrm{stimulus}$ leading to those
values respectively;
(vi)  the dynamic range is calculated as ${\it\Delta}=10
\log_{10}({\it p}_{0.9}/{\it  p}_{0.1})$.

Notice that in the active phase ${\it\rho}_{0.1}$ reaches a finite
steady state at exponentially large times in the limit
${\it\lambda}\to {\it\lambda}_c^+$. This makes the study of large
systems very lengthy in that parameter region.

Extended regions of enhanced response are found also by running our
simple dynamic protocols on the connectome network.  A way to
visualize the broadening of the critical region is presented in
Supplementary Figure S3, where the density of active sites given a
fixed active seed ${\it \rho}_{\textrm{f}}$ is plotted as a function of ${\it
  \lambda}$. The critical region broadens if compared to the case of a
regular (disorder-free) lattice of the same size.

{\bf Finite-size scaling.}  In the standard critical point scenario --
assuming the system sits exactly at the critical point but it runs
upon a finite system (of linear size L) -- the average density (order
parameter) starting from an initially active configurations decays as
${\it t}^{- \theta} \times \exp(-{\it t/\tau(L)})$, where the cut-off
time scales with system size, as ${\it\tau (L)n}\propto {\it L}^z$
(${\it z}$ the dynamic critical exponent), allowing us to perform
collapse plots for different system sizes.  Instead, in Griffiths
phases, the cut-off time does not have an algebraic dependence on
${\it L}$; it is the largest cluster which is cut-off by ${\it L}^d$,
and the corresponding escape/decay time from it grows like
${\it\tau(L)} \propto \exp({\it c L}^d)$. Therefore, even for
relatively small system sizes, such a cut-off is not observable in
(reasonable) computer simulations: order-parameter-decay plots should
exhibit power-law asymptotes without any apparent cut-off.  On the
other hand, the power-law exponent -- which can be estimated from a
saddle point approximation, dominated by the largest rare region -- is
severely affected by finite size effects.  Therefore, summing up,
while in standard critical points finite size effects maintain the
critical exponent but visibly affect the exponential cut-offs, in
Griffiths phases, apparent critical exponents are affected by
finite-size corrections while exponential cut-offs are not observable.
Unless otherwise specified, simulations on HMNs are for systems of
size ${\it N}=2^{14}=16384$. Supplementary Figure S4 shows that upon
increasing the system size (and the number of hierarchical levels
accordingly), the picture of generic power-law decay of activity
remains valid in the whole GP. One can observe that for each value of
${\it\lambda}$, the effective exponent tends to an asymptotic value,
which is expected to hold in the large-network-size limit.

{\bf Avalanches in the human connectome.}
Unlike avalanches on synthetic HMNs, typically run on several
($10^7-10^8$) network realizations, avalanche statistics on the human
connectome network is the result of a large number of avalanches
($>10^9$) on the unique network available. Such limitation explains
the emergence of strong cutoffs in the avalanche-size distributions
(see Figure 4). In the main text, we propose to fit avalanche size
distributions with truncated power laws, reflecting the superposition
of generic power-law behavior and finite-size effects. In order to
assess the validity of our hypothesis, we resort to the
Kolmogorov-Smirnov method, by which the best fit is provided by the
fitting function $g({\it S})$, which minimizes the estimator
\begin{equation}
{\it D}_\textrm{KS}=\max |G(S)-F(S)|,
\end{equation}
where $G({\it S})$ is the cumulative distribution associated with
$g({\it S})$ and $F({\it S})$ is the cumulative distribution of
empirical data (simulation results, in our case).

In case of limited amount of empirical data, the use of diverse
fitting techniques (least squares, maximum likelihood etc.) is
advised, in order to avoid biases.  However, given the abundance of
data in our case, a non-linear least-squares fit provides a reliable
estimate of parameters (note that a truncated power-law cannot be fit
linearly by standards methods). We recall that the least-squares
method is essentially a minimization problem: given a set of empirical
data points $({\it x}_{i},{\it y}_{i})_{i=1,\ldots n}$ and a fitting
function $g({\bf x}, {\bf y}, {\bf b})$ depending on a set of
parameters ${\bf b}$, the fit is provided by the set of parameters
that minimizes the function $\sum_{i=1}^n [{\it y}_i-g({\it x}_i,{\bf
  b})]^2$. In the case of a linear regression, such minimization can
be performed exactly. In the case of a non-linear fit, instead,
the minimization has to be performed numerically.

For every value of ${\it \lambda}$ (every curve in Figure 4), we
proceed as follows: (i) we provide a least-squares fit for the
avalanche-size distribution, based upon the truncated power-law
hypothesis and calculate the corresponding Kolmogorov-Smirnov ${\it
  D}_\textrm{KS}$; (ii) we repeat the above procedure for alternative
candidate distributions (non-truncated power law and exponential)
(iii) we compare the results for the ${\it D}_\textrm{KS}$ indicators
and choose the hypothesis with the smallest ${\it D}_\textrm{KS}$ as
the best fit.  In Supplementary Figure S5 we provide an example of
this procedure, as obtained from our data for ${\it\lambda}=0.017$. In
this case, as in every other case examined, the truncated power law
provides the best fit among the ones tested, both by the KS criterion
and upon visual inspection. Notice that also the power-law hypothesis
appears plausible to some extent, whereas the exponential hypothesis
deviates significantly from the data, however one chooses the minimum
avalanche size ${\it S}_\textrm{min}$ to fit.  Special attention has
been devoted to the choice of the lower bound ${\it S}_\textrm{min}$,
as advised in Ref \cite{Clauset}. Such a choice is usually made by
visual inspection for large systems, where it is easy to estimate
visually the point in which power-law behavior takes over. In small
systems, instead, a more quantitative procedure is required. For every
fit described above (points (i) and (ii)), we have chosen the best
estimate for the ${\it S}_\textrm{min}$ upon preliminarily applying a
KS procedure to different candidate values of ${\it S}$ (following
\cite{Clauset}). We found that in each case, the KS estimator
displayed a minimum for values of ${\it S}\approx 6$, for the
truncated power-law and power law hypotheses.

{\bf Spectral analysis.}  Let us call ${\it q}_i({\it t})$ the
probability that the node ${\it i}$ is active at time ${\it t}$.  The
density of active sites can be written as ${\it\rho(t)}=\langle {\it
  q}_i({\it t})\rangle$, averaged over the whole network. In the case
of Model B dynamics, the probabilities ${\it q}_i({\it t})$ obey the
evolution equation
\begin{equation}\label{eq:SIS}
\dot{{\it q}}_i({\it t})=-{\it q}_i({\it t})+{\it \lambda}[1- {\it q}_i({\it t})]
\sum_{i=1}^N {\it A}_{ij} {\it q}_j({\it t}),
\end{equation} 
where ${\it A}$ denotes the adjacency matrix and ${\it \lambda}$ the
spreading rate. Calling ${\it\Lambda}$ a generic eigenvalue of ${\it
  A}$, its corresponding eigenvector ${\bf f}({\it\Lambda})$ obeys
${\it A}{\bf f}({\it\Lambda})={\it\Lambda} {\bf
  f}({\it\Lambda})$. Working on undirected networks, all eigenvalues
${\it\Lambda}$ are real and any state of the system can be decomposed
as a linear combination of eigenvectors, as in
\begin{equation}\label{eq:decomposition}
{\it q}_i=\sum_\Lambda c({\it \Lambda}) f_i({\it\Lambda}).
\end{equation}
More importantly, if the network is connected (all our HMNs are),  the maximum
eigenvalue of {\it A}, ${\it\Lambda}_\textrm{max}$, is positive and unique (Perron-Frobenius
theorem, see e.g. \cite{gantmacher}).  As a consequence, it is commonly assumed
that the critical dynamics of Eq. (\ref{eq:SIS}) at ${\it\lambda}={\it\lambda}_\textrm{c}$ is
dominated by the leading eigenvalue ${\it\Lambda}_\textrm{max}$ and that, at the threshold
${\it\lambda}_\textrm{c}$, 
\begin{equation}\label{eq:approximation}
{\it q}_i\approx  c({\it \Lambda}_\textrm{max}) f_i({\it\Lambda}_\textrm{max}).
\end{equation}
Then one can impose the steady state condition $\dot{{\it q}}_i(t)=0$ and, under the
fundamental assumption of Eq. (\ref{eq:approximation}), derive the well known
result \cite{Goltsev12}
\begin{equation}\label{eq:lambda_c}
{\it \lambda}_\textrm{c}=1/{\it \Lambda}_\textrm{max}.
\end{equation}
This result relies on the
implicit assumption that the principal eigenvalue is significantly larger than
the following one. The existence of such spectral gap', separating
${\it\Lambda}_\textrm{max}$ from the continuum spectrum of {\it A}, is a quite common
feature in complex networks, being a measure of their {\it small-world}
property. However, the picture of cortical networks as hierarchical structures
distributed across several levels suggests that such systems may exhibit very
different properties. We will prove this in the following and show how the above
picture changes in HMNs.

Figure 5a shows the average eigenvalue spectrum of
the adjacency matrix ${\it A}$ for HMNs. A detailed analysis of the
peak structure is beyond the scope of this work. Notice the absence of
isolated eigenvalues at the higher spectral edge (see also Figure
5b). The principal eigenvalue ${\it \Lambda}_\textrm{max}$
is not clearly separated from the others. The spectral gap,
characterizing small-world networks, here is replaced by an
exponential tail of eigenvalues.
All such eigenvalues share a common feature: their corresponding
eigenvectors are localized, as shown in Figure
5c. All components are close to zero, except for a
few of them in each network, corresponding to a rare region of
adjacent nodes. We claim that such rare region are responsible for the
emergence of the Griffiths phase (GP) over a finite range of the
spreading rate ${\it \lambda}$.

Localization in networks can be measured through the inverse
participation ratio, defined as
\begin{equation}
\mbox{{\it IPR}}({\it\Lambda})=\sum_{i=1}^N f_i^4({\it\Lambda}).
\end{equation}
If eigenvectors are correctly normalized, $ \mbox{{\it
    IPR}}({\it\Lambda})$ is a finite constant of the order of $O(1)$
if $ {\it \Lambda}$ is localized, while $ \mbox{{\it
    IPR}}({\it\Lambda})\sim 1/\sqrt{{\it N}} $ otherwise.  Such
localization estimator is usually calculated for the principal
eigenvalue ${\it\Lambda}_\textrm{max}$ of a network. Indeed Figure
5d shows that $IPR$ is finite and insensitive to
changes in systems sizes. On the other hand, upon increasing the
density of inter-module connections, $IPR$ rapidly decreases,
suggesting that small-world effects enhance delocalization.  Having
introduced a criterion to identify localized eigenvectors, we found
that a whole range of eigenvalues below ${\it\Lambda}_\textrm{max}$
correspond to localized eigenvectors. The structure of
Eq. (\ref{eq:SIS}) and its solutions suggest that while a spreading
rate of the order of ${\it\lambda}=1/{\it\Lambda}_\textrm{max}$ allows the
system to access the localized behavior dictated by the eigenvector
${\bf f}({\it\Lambda}_\textrm{max})$, lager values of ${\it\lambda}$ grant
access to the next eigenvalues and eigenvectors that are less and less
localized.  This ultimately establishes a strong connection between
eigenvalue localization and rare-region effects.


 


 


\newpage

\section*{Acknowledgements} We acknowledge financial support from Junta de Andalucia, grant
  P09-FQM-4682.  We thank Olaf Sporns for kindly giving us access to the
  human connectome data. 

\section*{Author contributions}
PM and MAM designed the analyses, discussed the results, and wrote the
manuscript. PM wrote the codes and performed the simulations.
\newline
\newline
\noindent {\bf Competing financial interests}: The authors declare no competing financial interests.

\end{document}


\title{SUPPLEMENTARY INFORMATION:
Griffiths phases and the stretching of criticality in brain networks
}

\author{Paolo Moretti and Miguel A. Mu\~noz}

\textwidth = 470pt
\oddsidemargin = 0pt
\maketitle

\begin{figure}[h]
\begin{center}
  \includegraphics*[width=10cm]{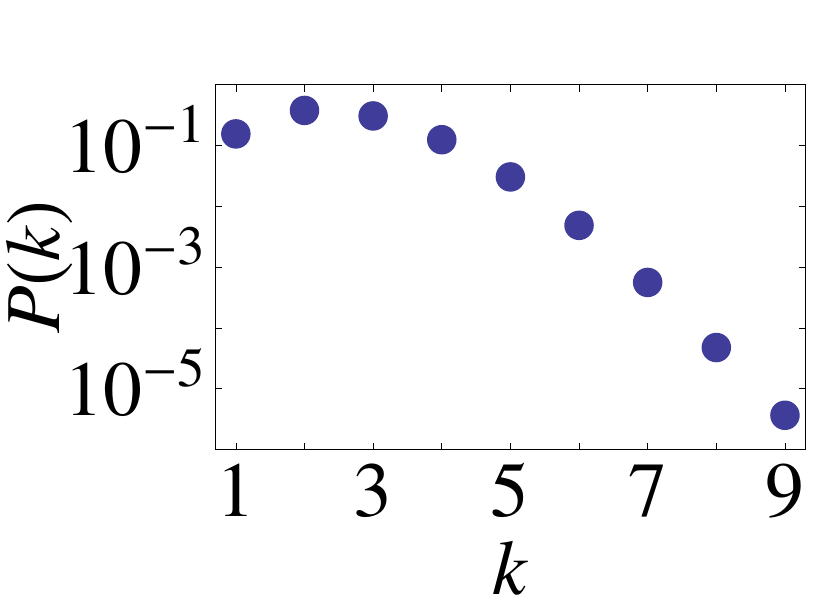}
  \caption{ {\bf Degree distribution of hierarchical networks.}  Average degree distribution for HMN-1 networks. In this case:
$N=2^{14}$, $s=13$, $M_0=2$, $p=1/4$, $\alpha=1$. Analogous distributions are
found for HMN-2 networks.} 
    \label{fig:Topology}
\end{center}
\end{figure}

\begin{figure}[t] 
\begin{center}
  \includegraphics*[width=15cm]{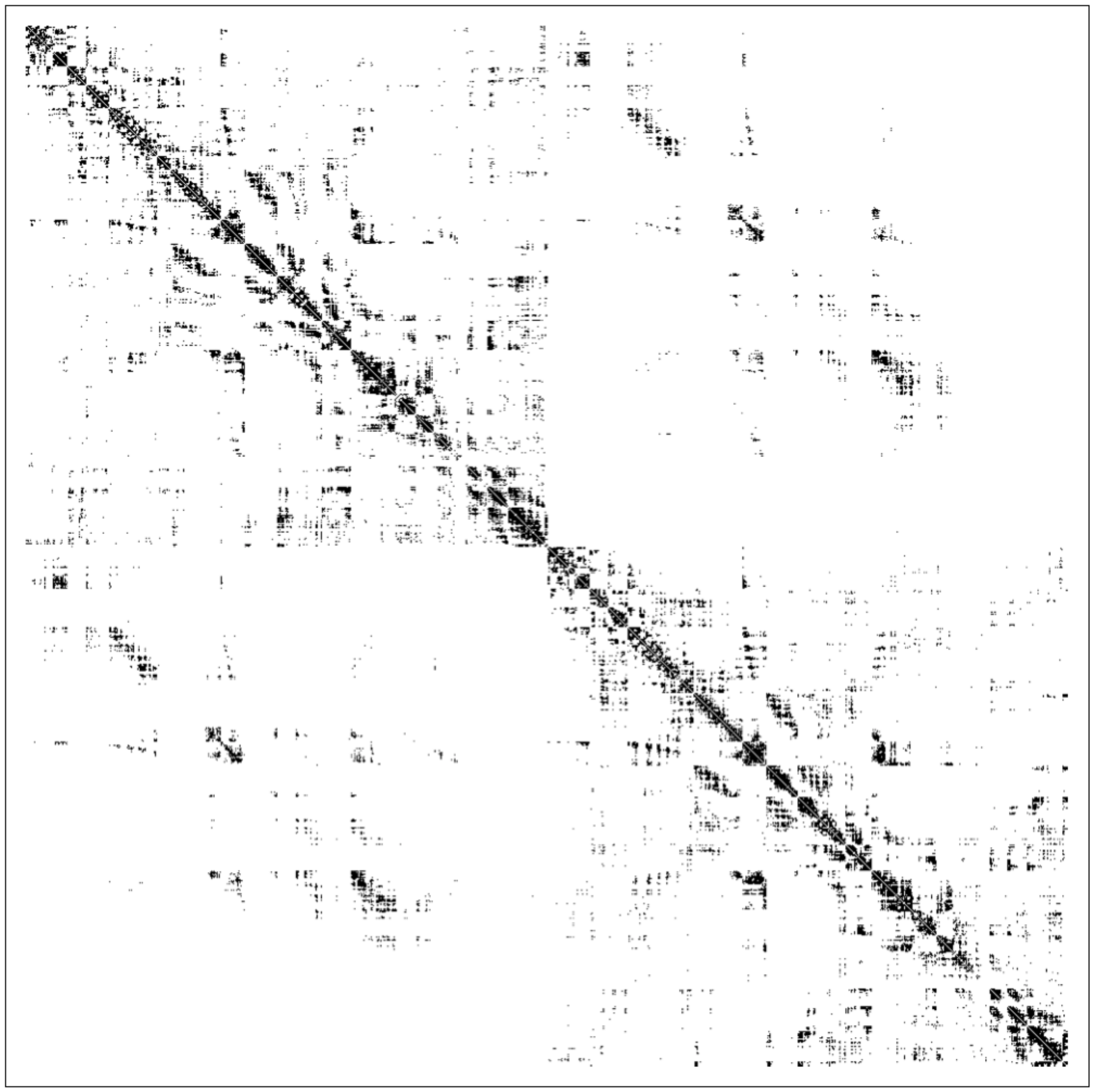}
  \caption{{\bf Hierarchical structure of the human connectome.} 
  An array plot of the adjacency matrix of the human connectome network is shown. A black square is
placed at every non-zero entry of the matrix. Even upon simple visual inspection, the network
appears characterized by a markedly hierarchical structure.}
    \label{fig:connectome_hierarchy}
\end{center}
\end{figure}

\begin{figure*}[h]
\begin{center}
  \includegraphics*[width=10cm]{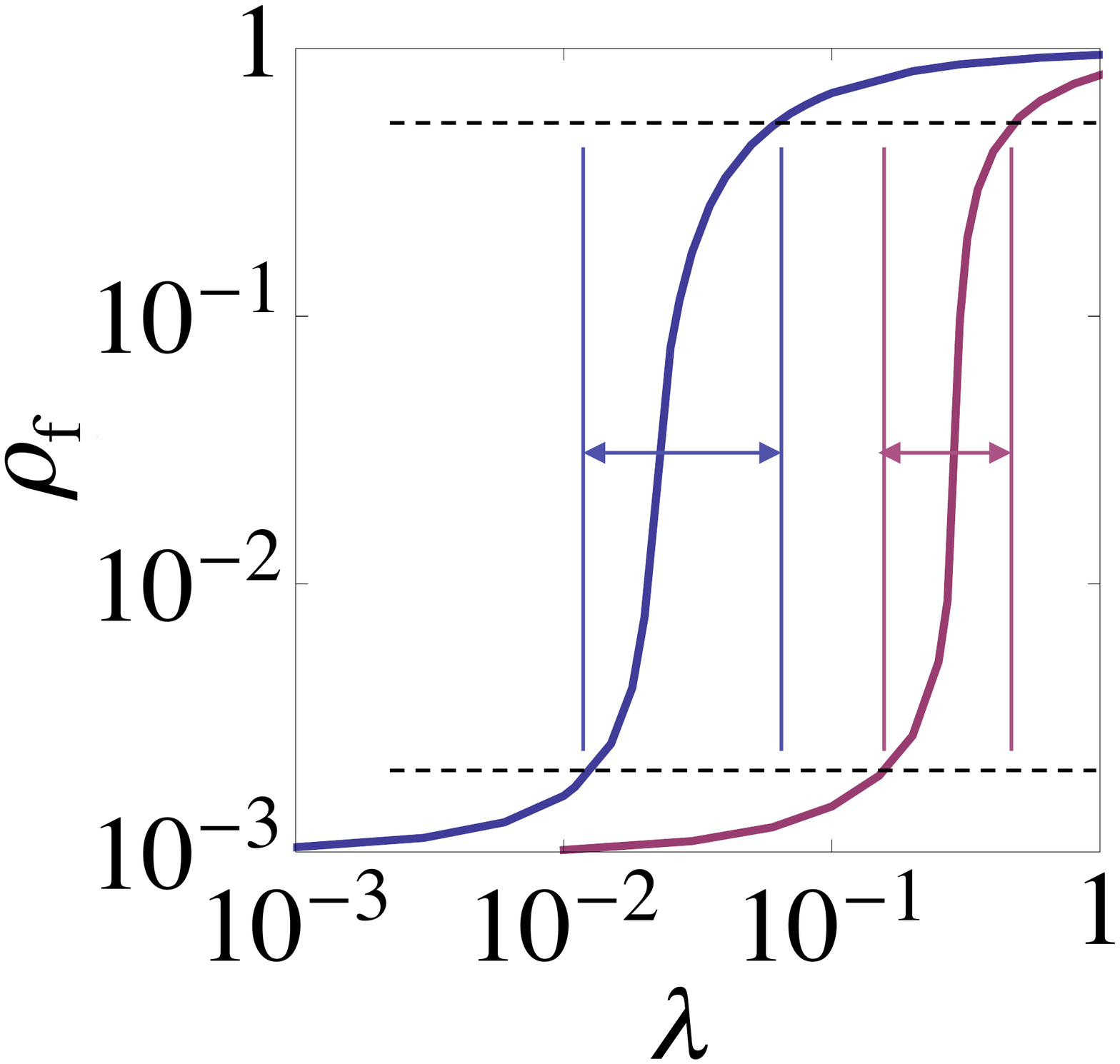}
  \caption{{\bf Extended region of activity in the human connectome network.} 
Activity in the human connectome (blue line), compared to a (disorder-free) square lattice of the
same size (red line). Data are obtained for Model B simulations. Cuts in the vertical axis (dashed
black lines) are located at arbitrary heights for visual purposes, showing that the ascent of activity at
the transition is significantly less steep for the connectome. Notice also that activity for the
connectome extends over a decade more than for the square lattice.}
    \label{fig:activity}
\end{center}
\end{figure*}

\begin{figure*}[h]
\begin{center}
  \includegraphics*[width=10cm]{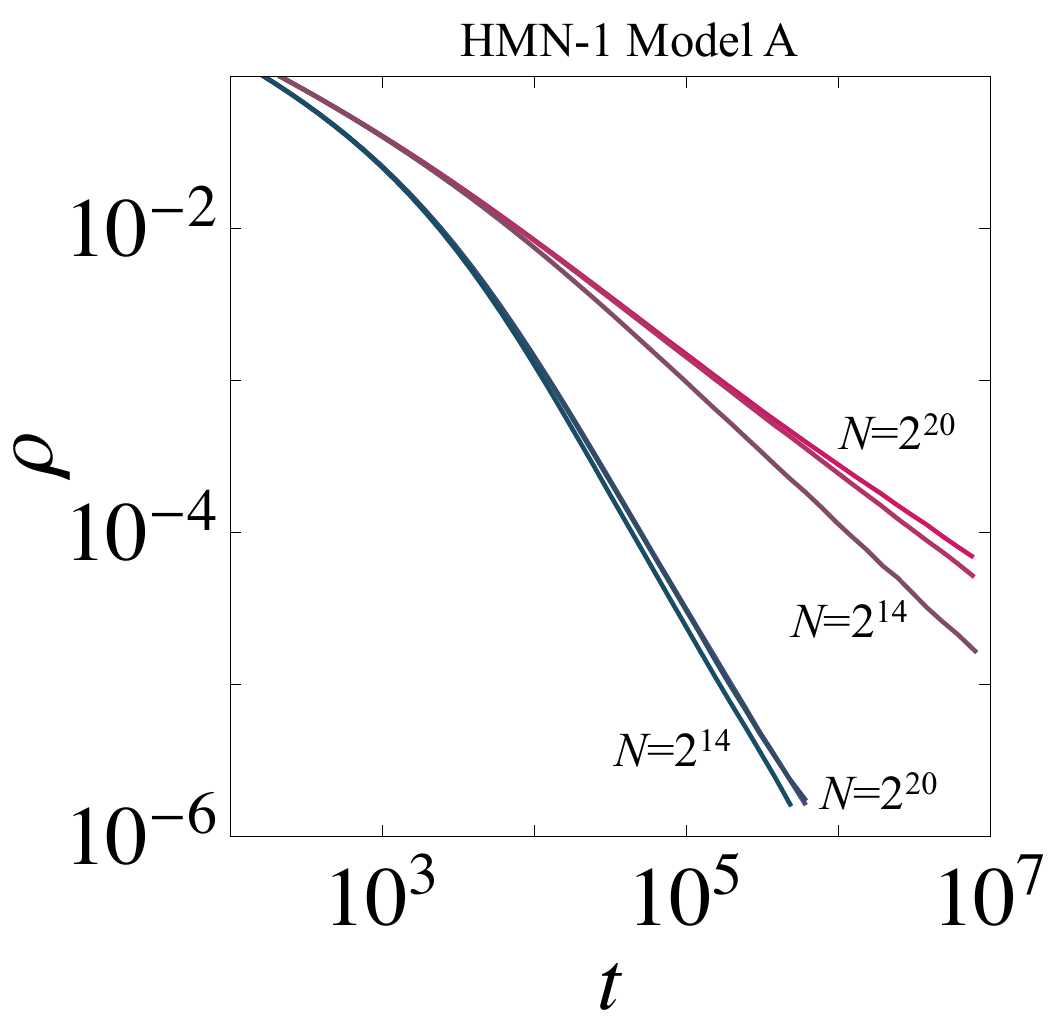}
  \caption{{\bf Finite-size scaling of activity decay in hierarchical networks.} 
HMN-1 networks, Model A, dynamic protocol (i). 
The density of active sites is plotted as a function
of time, for two different values of $\lambda$ within the GP: 
$\lambda=2.25$ (blue lines) and $\lambda=2.28$ (red lines). For
each value of $\lambda$, $\rho$ is plotted for $N=2^{14}$, $2^{17}$, 
and $2^{20}$ ($M_0=2$ in all cases), thus spanning the size range
$10^{4}$--$10^{6}$. The generic power-law behavior is conserved in the thermodynamic limit, as for each $\lambda$ the
slope of activity decay tends to an asymptotic limit. 
As explained in the text, finite size effects are quite 
different in GPs with respect to the standard critical point scenario.
}
\end{center}
\end{figure*}

\begin{figure*}[h]
\begin{center}
  \includegraphics*[width=10cm]{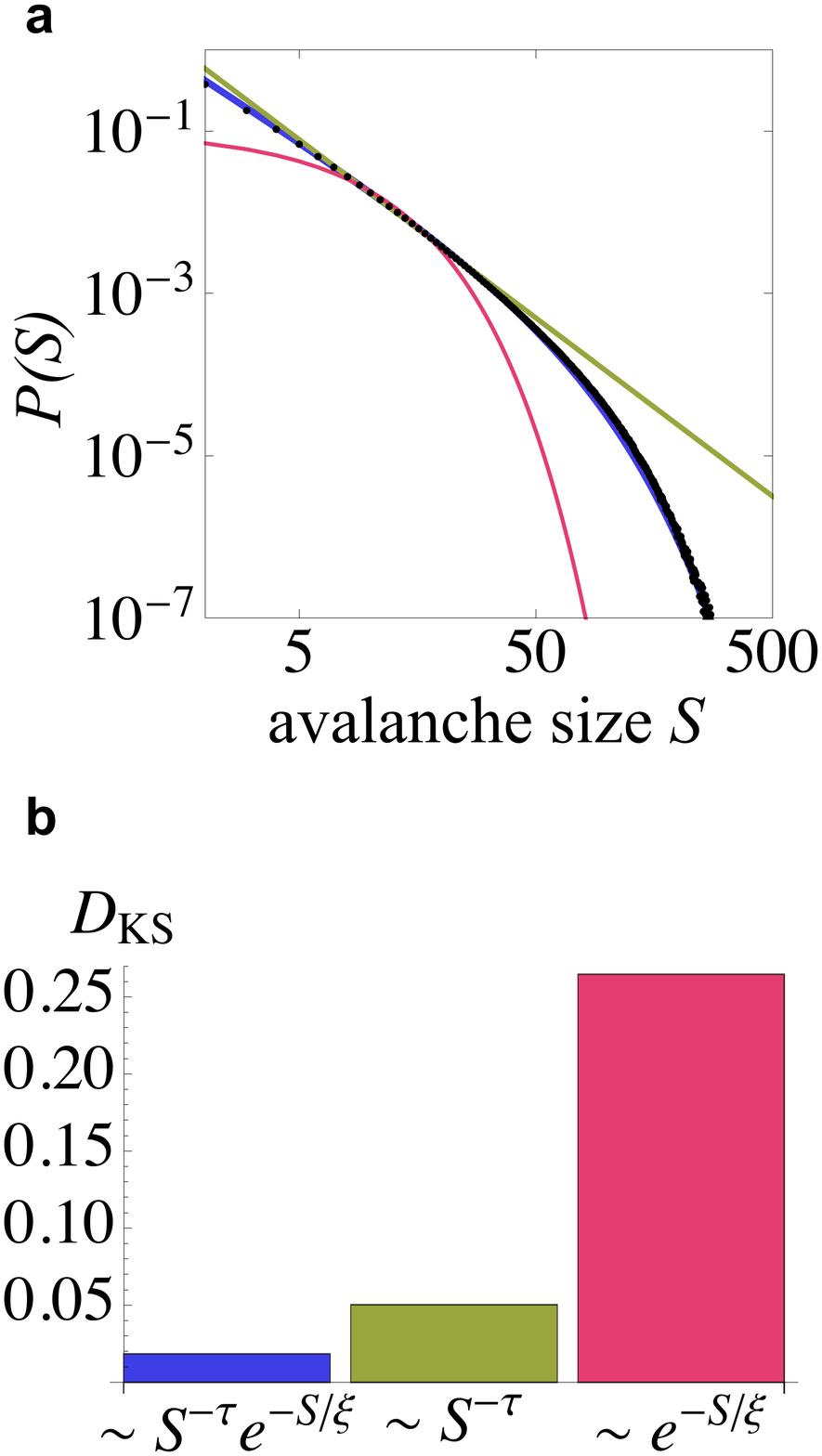}
  \caption{{\bf Analysis of avalanche-size distributions for the human connectome.} 
(a) Fit procedure for Model B avalanches in the human connectome network. The case of $\lambda=0.017$ is
displayed. Empirical simulation data (black dots) are compared to the best truncated-power-law fit
(blue line), power-law fit (green line) and exponential fit (red line). (b) The corresponding
Kolmogorov-Smirnov indicators are compared.
 }
    \label{fig:ks}
\end{center}
\end{figure*}